\documentclass[9pt]{extarticle}
\usepackage{spconf,amsmath,graphicx}

\def\etal{\emph{et al.~}}
\usepackage[table]{xcolor}
\usepackage{booktabs} % To thicken table lines
\definecolor{Pastel}{RGB}{250,250,255}
\definecolor{Pastel2}{gray}{0.99}

\usepackage[a4paper,bindingoffset=0.2in,%
left=0.5in,right=0.55in,top=0.86in,bottom=0.7in,%
footskip=.25in]{geometry}
\usepackage{graphicx} % graficos
\usepackage{multirow, array} % para las tablas
\usepackage{mathtools}
\usepackage{amssymb}

\usepackage{graphicx}

\newcommand{\elem}[3]{\left[#1\right]_{#2, #3}}

\newcommand{\bt}[1]{\mbox{$\bf #1$}}
\def\l{\left(}
\def\r{\right)}
\title{Image Coding via Perceptually Inspired Graph Learning}
\name{Samuel Fernández-Menduiña, Eduardo Pavez, and Antonio Ortega\thanks{Author email: samuelf9@usc.edu. SFM was funded in part by the Fulbright Commission in Spain.}}
\address{University of Southern California, Los Angeles, CA, USA}
\begin{document}
%\ninept
%
\maketitle
\begin{abstract}
Most codec designs rely on the mean squared error (MSE) as a fidelity metric in rate-distortion optimization, which allows to choose the optimal parameters in the transform domain but may fail to reflect perceptual quality. Alternative distortion metrics, such as the structural similarity index (SSIM), can be computed only pixel-wise, so they cannot be used directly for transform-domain bit allocation. Recently, the irregularity-aware graph Fourier transform (IAGFT) emerged as a means to include pixel-wise perceptual information in the transform design.  This paper extends this idea by also learning a graph (and corresponding transform) for sets of blocks that share similar perceptual characteristics and are observed to differ statistically, leading to different learned graphs. 
%to re-purpose graph learning algorithms considering perceptual criteria. 
We demonstrate the effectiveness of our method with both SSIM- and saliency-based criteria. We also propose a framework to derive separable transforms, including separable IAGFTs. An empirical evaluation based on the 5th CLIC dataset shows that our approach achieves improvements in terms of MS-SSIM  with respect to existing methods.
\end{abstract}
\section{Introduction}
\label{sec:intro}
%Mean squared error (MSE) remains popular as a fidelity criterion to design coding methods and is commonly integrated into state of the art encoders
%. Being easy to handle mathematically, it allows 
%for  direct optimization of the codec parameters. 

Mean squared error (MSE) remains popular as a fidelity criterion for codec design because it simplifies transform-domain parameter optimization, even though it often fails to reflect perceptual quality \cite{Girod1993}. 
In recent years, research has focused on studying alternatives to include  perceptually-driven metrics \cite{WangBSS2004, WangSB2003} in the coding pipeline. Some of these ideas have succeeded from a practical viewpoint, e.g., quantization tables that can better reflect the human visual system \cite{WangLC2001, Watson1993}. Nonetheless, in the theoretical front, including perceptual metrics as part of the rate-distortion optimization (RDO) used by most codecs remains an open problem.

In this sense, the main challenge is the widespread adoption of transform coding \cite{Goyal2001} for image and video compression. Designing these codecs becomes simpler when using the MSE: by virtue of Parseval's identity, the distortion is the same in the pixel and the transform domain, allowing to perform RDO in the transform domain. On the other hand, many widely used perceptual metrics (e.g., SSIM and its variations) can only be computed in the pixel domain. Moreover, these metrics usually require a complete decoded image to be computed; therefore, codec optimization becomes an iterative process based on evaluations \emph{a posteriori} (encode, decode, estimate the metric and repeat the process). For instance, Pergament \etal \cite{Pergament+2022} adjusted the quantization parameter (QP) of modern video compression systems, such as  HEVC, to account for perceptual importance. Yet, this method cannot work at the pixel level {and the mapping between perceptual importance and QP follows no optimality criterion}. Recently, deep-learning methods \cite{Balle+2020} have succeeded in optimizing perceptual metrics like the SSIM, 
but these models require millions of floating point operations per pixel and end-to-end training \cite{Guleryuz+2021}.

To address these challenges, Lu \etal \cite{LuOMC2020} designed transforms based on an inner product that gives different weights to different pixels. The corresponding weighted MSE (WMSE) can be designed so that the weights capture the relative perceptual importance of each pixel, as estimated by metrics such as SSIM. This work relied on graph signal processing (GSP) \cite{ShumanNFOV2013, OrtegaFKMV2018,Ortega2022} principles, %, which in the last decade has emerged as a powerful tool for transform coding. Besides generalizing existing algorithms, 
and in particular on the interpretation of transforms such as the discrete cosine transform (DCT) \cite{Strang1999} or the asymmetric discrete sine transform (ADST) \cite{HanSMR2011} as graph Fourier transforms (GFT) of specific path 
graphs. 
In particular, \cite{LuOMC2020} uses the irregularity-aware graph Fourier transform (IAGFT) to adapt the perceptual quality pixel-wise while using the DCT. The WMSE weights are chosen with the SSIM \cite{WangBSS2004} as a distortion metric after assuming an additive noise model for the transform coefficients. Experimental results {show improvements} in the multi-scale SSIM \cite{WangSB2003} with respect to JPEG. 

Despite its conceptual novelty, the method of \cite{LuOMC2020} has limitations. First, the authors chose the $2$D-DCT graph, modifying the transform by changing the definition of inner product via the IAGFT. Yet, the optimal graph may differ from the $2$D-DCT graph. In this sense, previous works in a different context \cite{FracastoroTF2019} suggest that accounting for the statistics of the image via graph learning \cite{EgilmezPO2018} leads to compression gains. Moreover, the solution of Lu \etal \cite{LuOMC2020} focuses on non-separable transforms, which are too complex for practical implementations. Also, the empirical validation is limited to a database of four images.

In this paper, we improve significantly on the performance of \cite{LuOMC2020} for perceptually inspired image coding by introducing a graph learning stage \cite{EgilmezPO2018, PavezOM2017} to design the transforms. We use the fact  that different perceptual regions of an image have different statistical properties \cite{SimoncelliO2001}; thus, the graphs modeling the correlation between pixels with each of these regions are different. We propose to use a perceptual metric to  classify the blocks of each image into several classes, so that different graphs can be learned within each class. To provide an empirical validation, we modify a JPEG encoder using {these perceptually} learned graphs for transform coding. 

Since our method is based on the properties of different perceptual regions, we can use a family of fixed transforms chosen based on perceptual criteria. These transforms are available at both the encoder and the decoder so the side information is the index for a set of predefined transforms, as in \cite{LuOMC2020}. As perceptual criteria, we consider both the SSIM and saliency \cite{IttiKN1998}. {Although some papers have already combined graph learning and compression  \cite{FracastoroTF2016, LiaoCMYH2018, EgilmezCO2020}, to the best of our knowledge, none of them include perceptual criteria in the estimation process}.

In a practical setup, low-complexity solutions are often preferred. Thus, we also show that perceptually inspired graph learning can be constrained to obtain separable transforms. To do so,  we estimate a covariance model based on combinatorial graph Laplacians  using bi-convex optimization \cite{PavezOM2017}, leading to a novel, learned separable IAGFT. Finally, we assess the performance of both the separable and the non-separable methods in a set of $30$ images from the $5$th CLIC dataset \cite{CLIC2022}. 
As distortion metric, we choose the MS-SSIM since 1) it aligns better with the human visual system \cite{WangSB2003} and 2) the method in \cite{LuOMC2020} obtains the best results with this quantity. Our algorithm using non-separable transforms requires $6$\% less bit rate on average than the method proposed in \cite{LuOMC2020}, while the method using separable transforms reduces the bit rate more than $5$\%. Introducing structural constraints reduces the performance of the system. However, for separable transforms, we obtain computational gains: they are $30$\% faster than their non-separable counterparts. 

\section{Preliminaries}
\textbf{Notation.} Uppercase bold letters, such as $\bt A$, denote
matrices. Lowercase bold letters, such as $\bt a$, denote vectors.
The $n$th component of the vector $\bt a$ is $a_n$, and the $(i, j)$th component of the matrix $\bt A$ is $a_{ij}$. We denote the vector of all-ones by $\bt 1$. Regular letters denote
scalar values. %Image blocks are square arrays of size $\sqrt{n}\times \sqrt{n}$. In $\bt A = \text{diag}(\bt b)$, $\bt A$ is a diagonal square matrix such that $a_{ii} = b_i$. We use $\otimes$ to denote the Kronecker product between two matrices.
\subsection{Graph Signal Processing}
Let $\mathcal{G} = (\mathcal{V}, \mathcal{E}, \bt W)$ denote a weighted undirected graph, where $\mathcal{V}$ is the vertex set, $\mathcal{E}$ is the edge set, and $\bt W$ denotes the weighted adjacency matrix. The entries of the matrix $\bt W$ satisfy $w_{ij}>0$; that is, every edge $(i, j)\in \mathcal{E}$ has positive weights. We define the combinatorial graph Laplacian (CGL) as $\bt L = \bt D - \bt W$, where $\bt D$ is the degree matrix such that $d_{ii} = \sum_{j = 1}^{n} \, w_{ij}$. We use the CGL to define the graph Fourier transform (GFT). Let $\bt L = \bt U \bt \Lambda \bt U^\top$ be the eigendecomposition of the CGL, where $\bt \Lambda$ is a diagonal matrix containing the eigenvalues and $\bt U$ is an orthonormal matrix with the corresponding unit norm eigenvectors. These eigenvectors are the basis for the GFT.

In a recent work, Girault \etal \cite{GiraultON2018} proposed the irregularity-aware graph Fourier transform (IAGFT). Under this framework, the Fourier modes are defined not only in terms of the signal variation operator but also introducing a positive definite matrix $\bt Q$ that leads to the $\bt Q$-inner product, $\langle \bt x, \bt y \rangle_{\tiny \bt Q} = \bt x^\top \bt Q \bt y$. This inner product induces a new definition of orthogonality: $\bt x$ and $\bt y$ are orthogonal if and only if $\bt x^\top \bt Q \bt y = \bt 0$. In our work, the matrix $\bt Q$ is diagonal; therefore, the energy of the signal is given by $\vert \vert \bt x\vert \vert^2_{\tiny \bt Q} = \langle \bt x, \bt x \rangle_{\tiny \bt Q} = \sum_{i\in \mathcal{V}}\, q_i x_i^2$, where $\bt Q = \text{diag}\l q_1, \hdots, q_n\r$. The $(\bt L, \bt Q)$-GFT basis vectors are the columns of $\bt U = \left[\bt u_1, \hdots, \bt u_n \right]$, which solve the generalized eigenvalue problem
\begin{equation}
\bt L\bt u_i = \lambda_i\, \bt Q \bt u_i, \, \text{where } \lambda_1 \leq \lambda_2 \leq \hdots \leq \lambda_n. 
\end{equation}
Note that the eigenvectors are $\bt Q$-orthonormal: $\bt U^\top \bt Q \bt U = \bt I$. We let $\bt F = \bt U^\top\bt Q$ be the forward transform and $\bt F^{-1} = \bt U$ be its inverse.
 
\subsection{Weighted MSE Optimization via IAGFT}
The IAGFT modifies the definition of graph signal energy, weighting the contribution of each node. In \cite{LuOMC2020}, the authors rely on this property to adapt the perceptual quality pixel-wise. Mathematically, let $\bt z$ be the reference signal and $\bt x$ be its distorted version. Also, let their IAGFTs be $\hat{\bt z}$ and $\hat{\bt x}$, respectively. Then, we can define the WMSE as
\begin{equation*}
    \text{WMSE}(\bt z, \bt x, \bt q) = \frac{1}{n}\sum_{i = 1}^n q_i( z_i- x_i)^2 = \frac{1}{n}\sum_{i = 1}^n(\hat{z}_i- \hat{x}_i)^2,
\end{equation*}
where the last equality follows from the Generalized Parseval's Theorem \cite{GiraultON2018}. Let us assume a uniform quantizer in the transform domain. With the GFT, the variance of the error in the pixel domain is uniform and equal to $\Delta^2/12$, where $\Delta$ is the quantization step. Nonetheless, the $\bt Q$-orthogonality of the IAGFT introduces a weighting factor, modifying the variance to $\Delta^2/(12q_i)$. From this model, the authors of \cite{LuOMC2020} derive the matrix $\bt Q$ that optimizes the SSIM \cite{WangBSS2004}:
\begin{equation}
\label{eq:q_ssim_rule}
    q_i = \frac{(n + \sum_{i = 1}^n \gamma_i)\sqrt{\gamma_i}}{\sum_{i = 1}^n \sqrt{\gamma_i}} - \gamma_i, 
\end{equation}
where $\gamma_i = \Delta^2 /(12(2\sigma^2_{x_i}+c_2))$, $\sigma^2_{x_i}$ is the local variance, and $\mathbf{Q}=\mathrm{diag}(q_1, \hdots, q_n)$.
\subsection{Graph learning}
\label{sec:ges}
In this work, we consider both separable and non-separable transforms. Regarding non-separable transforms, we follow the formulation in \cite{EgilmezPO2017}: to estimate the combinatorial graph Laplacian, we minimize the objective function
\begin{equation}
J(\bt L) = -\log \det (\bt L + \bt 1 \bt 1^\top /n) + \text{tr}(\bt L \bt S),
\end{equation}
Given a graph topology $\mathcal{E}$, we only consider matrices from the set of CGLs:
\begin{equation}
    \bt L = \lbrace \bt L \in \mathcal{S}_n^+: \bt L\bt 1 = \bt 0, \elem{\bt L}{i}{j} \leq 0 \text{ for } i \not = j \rbrace,
\end{equation}
where $\mathcal{S}_n^+$ is the set of semi-positive definite matrices of size $n\times n$. To solve the optimization problem, we follow the block-coordinate descent method proposed in \cite[Algorithm~2]{EgilmezPO2017}.

In a separable transform, for a given $\sqrt{n}\times \sqrt{n}$ image block $\bt X_b$, the transform coefficients are given by $\hat{\bt X}_b = \bt U_c^\top \bt X_b \bt U_r$ \cite{PavezOM2017}. Different design options are possible for finding $\bt U_c$ and $\bt U_r$. In this work, we rely on the bi-convex formulation by Pavez \etal \cite{PavezOM2017}: we assume that $\bt U_r$ and $\bt U_c$ are the eigenvector matrices of two generalized Laplacians, $\bt M_r$ and $\bt M_c$, respectively. Both matrices are $\sqrt{n}\times \sqrt{n}$. Then, by minimizing
\begin{equation}
\label{eq:learn_sep}
-\log\det(\bt M_r \otimes \bt M_c) + \text{tr}((\bt M_r \otimes \bt M_c)\bt S), 
\end{equation}
we can obtain $\bt M_r$ and $\bt M_c$. In essence, we follow an iterative approach fixing either $\bt M_c$ or $\bt M_r$ and optimizing the expression in terms of the other matrix. We refer the reader to \cite{PavezOM2017} for details. In Sec.~\ref{sec:sepIAGFT}, we will modify Eq.~\eqref{eq:learn_sep} to obtain CGLs.

%rely on  Graph Based Separable Transforms (GBSP) \cite{EgilmezCOLY2016}. In essence, we focus on finding two weighted line graphs that characterize  columns and rows. In practice, this framework requires estimating two covariance matrices, one for the rows and another for the columns. Then, we can apply the graph learning method presented above using line-graph constrains.

\begin{figure*}
\centering
\includegraphics[scale = 0.45]{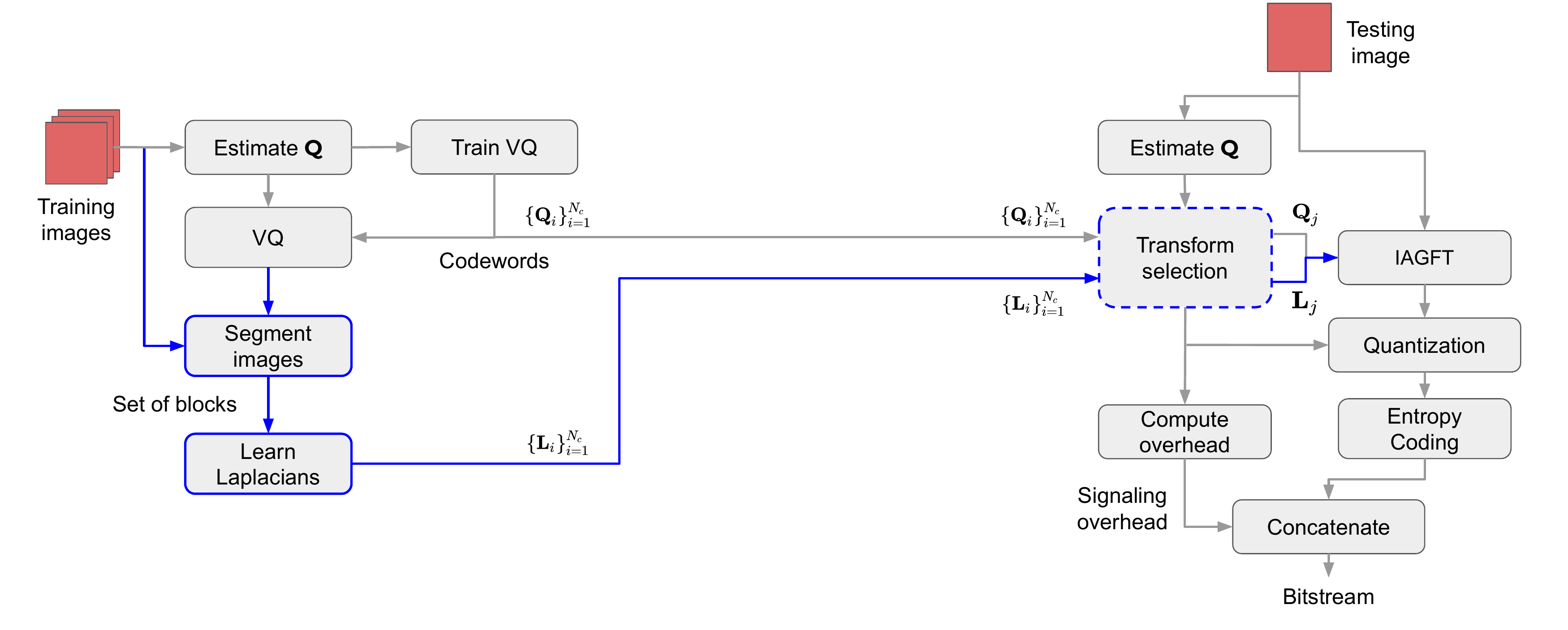}
%\includegraphics[scale = 0.50]{ICIP2023_BD_TOP.png}
%\hspace{8em}
%\includegraphics[scale = 0.40]{ICIP2023_BD_BOTTOM.png}
\caption{Block diagram of our algorithm. Blocks in blue are new components. Left: Training stage. Right: Modified JPEG encoder, including the CGLs and the $\bt Q$ learned in the training stage. By interchanging the IAGFT with the DCT, we recover the conventional JPEG encoder.}
\label{fig:blockdiag}
\end{figure*}

\section{Perceptual graph learning}
Perceptual importance is linked to the statistical properties of an image \cite{SimoncelliO2001}. For instance, the human visual system is less sensitive to noise in textured regions, which contain mainly high-frequency information. In such regions, pixels are barely correlated to their neighbors. From a GSP perspective, correlation between pixels can be modeled using a graph. Therefore, regions with different levels of perceptual importance should be modeled using different graphs. 

In this section, we leverage this property to propose transforms for image compression.  To do so, first we show in Sec.~\ref{sec:pis} how to segment an image to obtain different perceptual regions. Then, in Sec.~\ref{sec:gl}, we discuss the graph-learning process, while in Sec.~\ref{sec:sepIAGFT}, we present our separable IAGFT. 

\subsection{Perceptual image segmentation}
\label{sec:pis}
We start from a set of $T$ training images, $\lbrace \bt X_t\rbrace_{t = 1}^T$. Although size may differ, for simplicity, we assume all of them are $M\times M$ arrays. We let the vectorized version of these images be $\bt x_t \in \mathbb{R}^{M^2}$ for $t = 1, \hdots, T$. For each of these vectors, we compute the perceptual weights $\bt q_t\in\mathbb{R}^{M^2}$ using, for instance, Eq.~\eqref{eq:q_ssim_rule}. The computation of these weights involves the whole image, but most codecs compute transforms block-wise. Therefore, we need to identify the perceptual weights associated with each block. Assume we have $N_b$ blocks in each image. Then, we associate to the $k$th (vectorized) block of each training image $\bt x_{t, k}\in \mathbb{R}^{n}$ the corresponding weights $\bt q_{t, k}\in\mathbb{R}^{n}$ from $\bt q_t$, for $k = 1, \hdots, N_b$ and $t = 1, \hdots, T$.

In a practical setup, coding the perceptual weights of the image to compress requires an excessive overhead. In \cite{LuOMC2020}, the authors train a vector quantizer via an entropy-constrained vector quantizer (ECVQ) \cite{GershoG2012}, although simpler algorithms like $k$-means perform correctly. We denote as $N_c$ the number of codewords after training, and let $\lbrace \bt q^{(i)} \rbrace_{i = 1}^{N_c}$ be these codewords. Typically, codewords represent  different levels of perceptual importance.

Now, we leverage this information to segment the training images perceptually. For each $\bt q_{t, k}$ in the training set, we find the closest codeword in our dictionary, $\tilde{\bt q}_{t, k} \in \lbrace \bt q^{(i)}\rbrace_{i = 1}^{N_c}$. By doing so, we are assigning a perceptually-inspired label to each block of the training set $\bt x_{t, k}$. That is, we are able to segment the training images perceptually by linking each block with one of the $N_c$ codewords. In the following, we denote as $i_{t, k}$ the class assigned to each block.

%We provide an example for a particular image in Fig.~\ref{fig:perceptual_segm}. In the next section, we explain how to perform graph learning using this particular information.
\subsection{Graph learning stage}
\label{sec:gl}
Once we have segmented the image perceptually, we can perform graph learning on each category. The first step is to compute the covariance matrix. Define the scalar $m_{t, k}$ as the average intensity of the block $\bt x_{t, k}$. We approximate the expectation of each block by $\bt m_{t, k} =  m_{t, k}\bt 1$.

Including the perceptual information we obtained in the previous section, we can write the covariance estimator for the $i$th class as
\begin{equation*}
\bt S_i = \frac{1}{K_i-1}\sum_{t = 1}^{T}\sum_{k = 1}^{N_b} I(i_{t, k} = i)(\bt x_{t, k} - \bt m_{t, k})(\bt x_{t, k} - \bt m_{t, k})^\top,
\end{equation*}
where $K_i = \sum_{t = 1}^{T}\sum_{k = 1}^{N_b} I(i_{t, k} = i)$ is the number of blocks in the training set for each class and $I(\cdot)$ is the indicator function.

By using the methods we described in Sec.~\ref{sec:ges}, we can compute $N_c$ graph Laplacians, $\lbrace \bt L_i\rbrace_{i = 1}^{N_c}$. Given these Laplacians, we obtain the basis of the $(\bt L_i, \bt Q_i)$-GFT by setting $\bt Q_i = \text{diag}(\bt q^{(i)})$. By making these bases available for both the encoder and the decoder, we can tailor the transforms to each perceptual region.

\subsection{Separable IAGFT design}
\label{sec:sepIAGFT}
For image compression, transforms with a DC component usually perform better. However, the separable learning method proposed on Eq.~\eqref{eq:learn_sep} cannot handle CGLs. To circumvent this issue, we modify the learning problem as follows. Let $\bt M = (\bt M_r+\bt J) \otimes (\bt M_c+\bt J)$. Then, we can minimize
\begin{equation}
\label{eq:sep_opt}
 -\log\det(\bt M) + \text{tr}(\bt M\bt S),
\end{equation}
where $\bt J = \bt 1 \bt 1^\top /\sqrt{n}$, for all $\bt M_c$ and $\bt M_r$ in the set of CGL. This expression has a probabilistic interpetation in terms of Gauss-Markov random processes \cite{EgilmezPO2017}. The problem is still bi-convex; therefore, we can solve it using conventional optimization routines. In particular, it can be shown that, keeping $\bt M_r$ fixed, Eq.~\eqref{eq:sep_opt} simplifies to
\begin{equation}
-\log\det(\bt M_c) + \sum_{i = 1}^n \text{tr}((\bt M_r+\bt J)\bt x_i)\text{tr}(\bt M_c \bt x_i^\top),
\end{equation}
that is, a weighted version of the optimization problem for non-separable graphs, which is easier to optimize than Eq.~\eqref{eq:sep_opt}. The equation keeping $\bt M_c$ fixed is analogous. 

Although the formulation above is sufficient for the conventional GFT, extending the same idea to the IAGFT is not straightforward. One option is applying the transforms row- and column-wise using the appropriate perceptual weights. This approach requires a different transform for each column and each row. As an alternative, if the perceptual matrix has structure, we can use only one transform for all columns and one transform for all rows.

Let $\bt Q_b\in\mathbb{R}^{\sqrt{n}\times \sqrt{n}}$ be the weight matrix in block form, where $\bt q_b = \text{vec}(\bt Q_b)$. Assume $\bt Q_b = \bt q_r \bt q_c^\top$ and $\bt M = \bt M_r \otimes \bt M_c$. Let us define $\bt Q_c = \text{diag}(\bt q_c)$ and $\bt Q_r = \text{diag}(\bt q_r)$. Then, the eigenvector matrix $\bt U$ satisfying
\begin{equation}
\bt M \bt u_i = \lambda \, \text{diag}(\bt q_b)\bt u_i, \quad \text{ for } i = 1, \hdots, n,
\end{equation}
decomposes as $\bt U = \bt U_r \otimes \bt U_c$, where $\bt U_r$ is the basis for the $(\bt M_r, \bt Q_r)$-GFT and $\bt U_c$ is the basis for the $(\bt M_c, \bt Q_c)$-GFT. Therefore, we can use our conventional definition of separable transform, $\hat{\bt X} = \bt U_c^\top\bt Q_c \bt X \bt Q_r\bt U_r$.

In practice, we have to approximate the matrix $\bt Q_b$ by a rank $1$ matrix. We can do this by relying on the Eckart–Young theorem \cite{GolubL2013}. To assess both options, in Sec.~\ref{sec:exp_ssim} we will evaluate each of them.

\section{Empirical Evaluation}
We used images from the $5$th Challenge on Learned Image Compression dataset \cite{CLIC2022}. Since the database has no training set, we trained the vector quantizer and estimated the graphs with the thirty images in the validation set. For each of the thirty images in the testing set, we compute the perceptual weights, quantize them and identify the class of each block. Our encoder is depicted in Fig.~\ref{fig:blockdiag}. 

All images are cropped to $1024\times 1024$ pixels and converted to gray-scale before compression. We use two codewords in the vector quantizer \cite{GershoG2012} since the method in \cite{LuOMC2020} reaches its best performance with this configuration.

\subsection{Computational Complexity}
We consider only the complexity of the codec since the training step is performed only once. First, we need to compute and quantize the perceptual matrix $\bt Q$. This operation adds a constant overhead. Now, we distinguish between separable and non-separable transforms. For non-separable transforms, the complexity of the method is dominated by a matrix-vector product of size $n$. Therefore, we need $\mathcal{O}(n^2)$ products. For separable transforms, we need two matrix-matrix products of size $\sqrt n$; therefore the complexity is $\mathcal{O}(n^{3/2})$. %With respect to the conventional JPEG algorithm, the main restriction is the lack of specialized algorithms to compute the transform, which are available for the DCT.

\begin{figure*}[t]
\centering
\includegraphics[scale = 0.37]{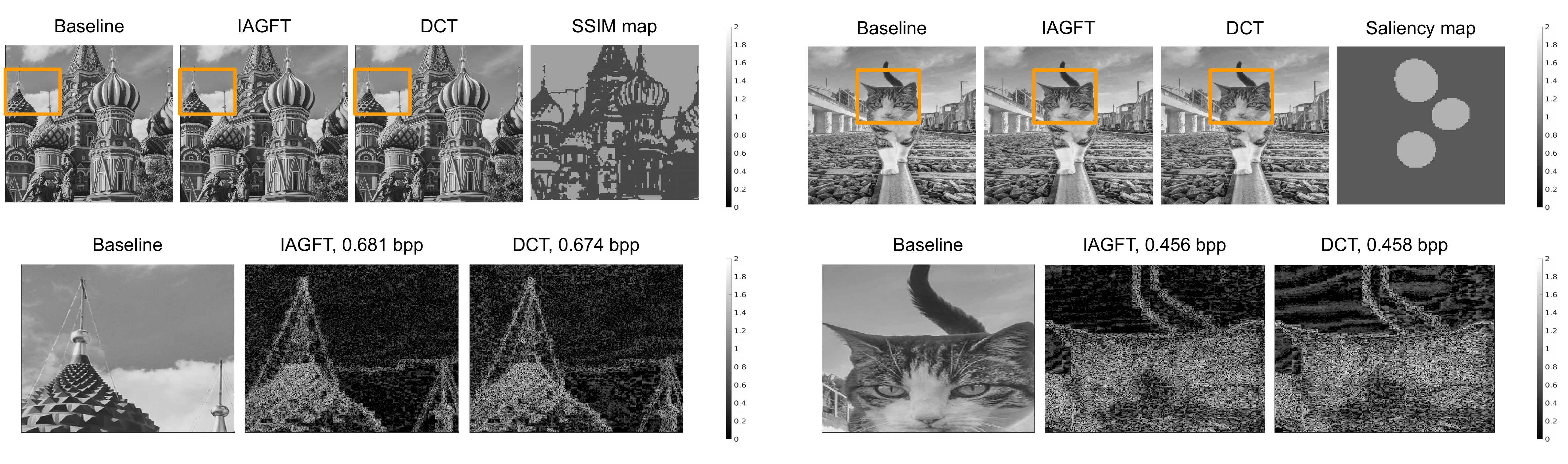}
\caption{Visual example using (left) SSIM and (right) saliency. Top: Original image, IAGFT image, DCT image, and pixel-wise perceptual weight map. Bottom: Detail of the original image and logarithmic magnitude of the compression error for both the IAGFT and the DCT. Using the IAGFT reduces the compression error in the area where the perceptual weight map takes higher values. This effect is apparent in the quantization error in the sky for the SSIM and in the region surrounding the cat's tail for saliency.}
\label{fig:perceptual}
\end{figure*}

%\begin{figure}[t]
%\centering
%\includegraphics[scale = 0.35]{saliency_perceptual_example.pdf}
%\caption{Visual example using the saliency. Top: Original image, JPEG image, IAGFT image, and saliency map. Bottom: Detail of the original image and logarithmic compression error for both the DCT and the IAGFT. Using the IAGFT reduces the compression error in the area where the saliency map takes higher values; this effect is apparent in the region surrounding the cat's tail.}
%\label{fig:saliency_perceptual}
%\end{figure}

The average computational time for JPEG in our workstation is $4.841$ s. Using our proposed algorithm, the time is $18.695$ s for non-separable and $13.021$ s for separable transforms. 

\subsection{SSIM-inspired transforms}
\label{sec:exp_ssim}
\begin{figure}
\centering
\includegraphics[scale = 0.265]{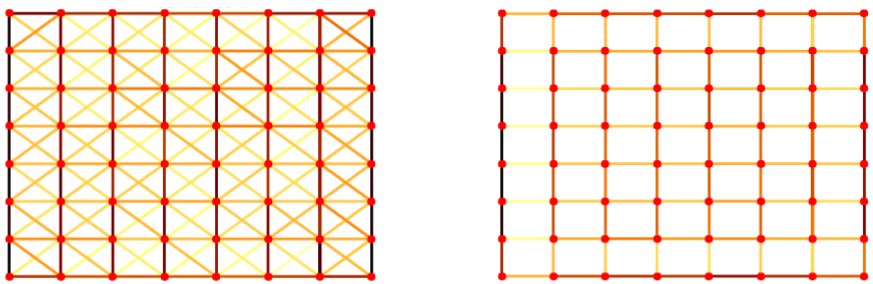}
\caption{Graphs learned using the SSIM and all connected constraints. We show blocks with high (left) and low (right) perceptual importance. Correlation is higher in perceptually important regions.}
\label{fig:graphs}
\end{figure}

In this section, we use the SSIM as our perceptual criterion; following the results from \cite{LuOMC2020}, we compute the optimal weights using the expression in Eq.~\eqref{eq:q_ssim_rule}. First, we show the graphs learned for each perceptual class using fully connected constraints in Fig.~\ref{fig:graphs}. %As we argued before, the graph corresponding to the higher level of perceptual importance has larger edge weights between closer nodes. Low-frequency information dominates perceptually important regions, which translates to higher correlations between pixels nearby.

We compress every image in the test set with our encoder. As in \cite{LuOMC2020}, we modify the quantization table by projecting it into the IAGFT basis; also, we use a run-length encoder to compress the perceptual weights. Then, we compute the Bjontegaard bit rate delta with respect to conventional JPEG using the MS-SSIM as distortion metric. We chose the MS-SSIM because 1) in practice, the MS-SSIM correlates better with the human visual system \cite{WangSB2003} and 2) the method proposed in \cite{LuOMC2020} obtains the best results with this quantity.

For non-separable transforms, we test three structural constraints: fully connected graphs, $8$ connected graphs, and $4$ connected graphs. Results are summarized in Table~\ref{tab:bj_results}, where we also included the method proposed in \cite{LuOMC2020} (IAGFT via the $2$-D DCT graph) and a modified version of our encoder using the conventional GFT with the perceptually learned graphs instead of the IAGFT. 

Using the fully-connected constraint leads to lower bit-rates on average, but the standard deviation of the results decreases when using more restrictive constraints since we have less parameters to estimate. Also, using the conventional GFT with the learned Laplacians reduces the performance of the encoder; we expected this result, since the GFT does not account for the differences on perceptual importance. In any case, this method still performs better than the algorithm in \cite{LuOMC2020}. We also provide a visual example in Fig.~\ref{fig:perceptual} (left).

\iffalse
\begin{figure}
\centering
\includegraphics[scale = 0.32]{perceptual_SSIM.pdf}
\caption{Perceptual example using the SSIM to compute $\bt Q$, jointly with the error between the compressed and the original image in logarithmic scale. Where $\bt Q$ is higher (sky, clouds) blocking artifacts are less severe.}
\label{fig:perceptual_ssim}
\end{figure}
\fi

\begin{table}[!t]
	\renewcommand{\arraystretch}{1.1}
	\small
	\centering 
	\begin{tabular}{ccccc}
		\toprule
		\textbf{Method} &  {Avg.} (\%) & STD & Min. (\%)  & Max. (\%)\\ 
		\midrule
		\rowcolor{Pastel2}
		  IAGFT $2$D-DCT \cite{LuOMC2020} &  $-0.584$ & $3.549$ & $ -8.483 $ & $12.766$ \\ 
		\rowcolor{Pastel}
		GFT, L (F) & $-0.271$ & $2.618$ & $ -3.772 $ & $6.096$ \\
        \rowcolor{Pastel2}
		IAGFT, L (F) & $\mathbf{-6.442}$ & $2.904$ & $ -12.183 $ & $-0.568$\\
 		\rowcolor{Pastel}
		IAGFT, L ($8$) & $-6.181$ & $2.705$  & $ -12.069$ & $-1.099$ \\
 		\rowcolor{Pastel2}
		IAGFT, L ($4$) & $-5.998$ & ${2.672}$  & $ -11.973 $ & $-1.659$\\ 
        \rowcolor{Pastel2}
        \midrule
		  IAGFT $2$D-DCT (SRC) &  $-3.391$ & $2.643$ & $ -8.591 $ & $1.924$ \\ 
        \rowcolor{Pastel}		
        GFT, L (SRC) & $-0.553$ & $1.077$ & $ -2.697 $ & $1.982$ \\ 
 		\rowcolor{Pastel2}
		IAGFT, L (SRC) & {$-5.732$ } & $2.596$  & $ -10.682 $ & {$-1.029$}\\	
        IAGFT, L (SR$1$) & $\mathbf{-5.876}$ & $2.662$  & $ -10.952 $ & {$-1.137$}  \\
		\bottomrule
	\end{tabular}
 	\caption{Bjontegaard delta bit-rate saving \cite{Bjontegaard2001} with respect to DCT-JPEG using the MS-SSIM  \cite{WangSB2003}. Negative numbers are compression gains. ``L'' indicates ``Learned Laplacian''. Structural constrains: all connected (F), $8$ connected ($8$), $4$ connected ($4$), row-column separable transform (SRC), and rank-$1$ separable transform (SR$1$). The IAGFT via the learned Laplacian minimizes the bit rate for separable and non-separable transforms (boldface).}
	\label{tab:bj_results}
\end{table}

Regarding separable transforms, we focus on the  learning algorithm in Sec.~\ref{sec:ges} (three last rows in Table~\ref{tab:bj_results}). The results follow the same trend. We can confirm that separability decreases the performance of the codec. Also, the rank $1$ decomposition offers better results than the row-column approach.

\subsection{Saliency-inspired transforms}
In this section, we design the perceptual weights to include information about saliency \cite{IttiKN1998}. Optimizing analytically the perceptual weights in terms of the saliency is intractable, so we resorted to a heuristic rule. We evaluated the expression in Eq.~\eqref{eq:q_ssim_rule} using the saliency map for the image \cite{WaltherK2006} instead of $\sigma_x^2$. We also set the constant $c_2 = 4$. In some cases, the saliency map is too localized; to increase the span of the salient regions, we smooth the saliency map using a Gaussian kernel. We provide an example in Fig.~\ref{fig:perceptual} (right).

%To compute $\bt Q$, we can use any perceptual measure. In this sense, analyzing the method in \cite{Pergament+2022} via $\bt Q$ instead of the quantization parameter (QP) is a promising venue for future research.

\section{Conclusion}
In this paper, we considered the problem of perceptual graph learning for image compression. Our idea is based on the fact that regions with different perceptual importance often differ statistically, so the graphs modeling the relationships between pixels on each of these regions are also different. Then, we can segment an image using perceptual criteria, like the SSIM, and apply graph learning methods on each region separately. By relying on this property, we can design perceptual transforms that adapt to the statistics of particular images. We also consider separable transforms and different perceptual criteria, such as saliency. When integrated into a JPEG codec, experiments in a large database show that our method requires $6$ \% less bit rate for a fixed MS-SSIM than the algorithm in \cite{LuOMC2020}. As future work, we plan to investigate how to add perceptual weights directly during the optimization process.

\bibliographystyle{IEEEbib}
\bibliography{IEEEabrv,PIT}

\begin{thebibliography}{10}

\bibitem{Girod1993}
Bernd Girod,
\newblock ``What's wrong with mean-squared error?,''
\newblock {\em Digital images and human vision}, pp. 207--220, 1993.

\bibitem{WangBSS2004}
Zhou Wang, Alan~C Bovik, Hamid~R Sheikh, and Eero~P Simoncelli,
\newblock ``Image quality assessment: from error visibility to structural
  similarity,''
\newblock {\em IEEE Transactions on Image Processing}, vol. 13, no. 4, pp.
  600--612, 2004.

\bibitem{WangSB2003}
Zhou Wang, Eero~P Simoncelli, and Alan~C Bovik,
\newblock ``Multiscale structural similarity for image quality assessment,''
\newblock in {\em The Thirty-Seventh Asilomar Conference on Signals, Systems \&
  Computers, 2003}. Ieee, 2003, vol.~2, pp. 1398--1402.

\bibitem{WangLC2001}
Ching-Yang Wang, Shiuh-Ming Lee, and Long-Wen Chang,
\newblock ``Designing {JPEG} quantization tables based on human visual
  system,''
\newblock {\em Signal Processing: Image Communication}, vol. 16, no. 5, pp.
  501--506, 2001.

\bibitem{Watson1993}
Andrew~B Watson,
\newblock ``{DCTune}: A technique for visual optimization of {DCT} quantization
  matrices for individual images,''
\newblock in {\em Sid International Symposium Digest of Technical Papers}.
  Citeseer, 1993, vol.~24, pp. 946--946.

\bibitem{Goyal2001}
Vivek~K. Goyal,
\newblock ``Theoretical foundations of transform coding,''
\newblock {\em IEEE Signal Processing Magazine}, vol. 18, no. 5, pp. 9--21,
  2001.

\bibitem{Pergament+2022}
Evgenya Pergament, Pulkit Tandon, Oren Rippel, Lubomir Bourdev, Alexander~G
  Anderson, Bruno Olshausen, Tsachy Weissman, Sachin Katti, and Kedar
  Tatwawadi,
\newblock ``{PIM}: Video coding using perceptual importance maps,''
\newblock {\em arXiv preprint arXiv:2212.10674}, 2022.

\bibitem{Balle+2020}
Johannes Ball{\'e}, Philip~A Chou, David Minnen, Saurabh Singh, Nick Johnston,
  Eirikur Agustsson, Sung~Jin Hwang, and George Toderici,
\newblock ``Nonlinear transform coding,''
\newblock {\em IEEE Journal of Selected Topics in Signal Processing}, vol. 15,
  no. 2, pp. 339--353, 2020.

\bibitem{Guleryuz+2021}
Onur~G Guleryuz, Philip~A Chou, Hugues Hoppe, Danhang Tang, Ruofei Du, Philip
  Davidson, and Sean Fanello,
\newblock ``Sandwiched image compression: wrapping neural networks around a
  standard codec,''
\newblock in {\em 2021 IEEE International Conference on Image Processing
  (ICIP)}. IEEE, 2021, pp. 3757--3761.

\bibitem{LuOMC2020}
Keng-Shih Lu, Antonio Ortega, Debargha Mukherjee, and Yue Chen,
\newblock ``Perceptually inspired weighted {MSE} optimization using
  irregularity-aware graph {Fourier} transform,''
\newblock in {\em 2020 IEEE International Conference on Image Processing
  (ICIP)}, 2020, pp. 3384--3388.

\bibitem{ShumanNFOV2013}
David~I Shuman, Sunil~K Narang, Pascal Frossard, Antonio Ortega, and Pierre
  Vandergheynst,
\newblock ``The emerging field of signal processing on graphs: Extending
  high-dimensional data analysis to networks and other irregular domains,''
\newblock {\em IEEE Signal Processing Magazine}, vol. 30, no. 3, pp. 83--98,
  2013.

\bibitem{OrtegaFKMV2018}
Antonio Ortega, Pascal Frossard, Jelena Kova{\v{c}}evi{\'c}, Jos{\'e}~MF Moura,
  and Pierre Vandergheynst,
\newblock ``Graph signal processing: Overview, challenges, and applications,''
\newblock {\em Proceedings of the IEEE}, vol. 106, no. 5, pp. 808--828, 2018.

\bibitem{Ortega2022}
Antonio Ortega,
\newblock {\em Introduction to graph signal processing},
\newblock Cambridge University Press, 2022.

\bibitem{Strang1999}
Gilbert Strang,
\newblock ``The discrete cosine transform,''
\newblock {\em SIAM review}, vol. 41, no. 1, pp. 135--147, 1999.

\bibitem{HanSMR2011}
Jingning Han, Ankur Saxena, Vinay Melkote, and Kenneth Rose,
\newblock ``Jointly optimized spatial prediction and block transform for video
  and image coding,''
\newblock {\em IEEE Transactions on Image Processing}, vol. 21, no. 4, pp.
  1874--1884, 2011.

\bibitem{FracastoroTF2019}
Giulia Fracastoro, Dorina Thanou, and Pascal Frossard,
\newblock ``Graph transform optimization with application to image
  compression,''
\newblock {\em IEEE Transactions on Image Processing}, vol. 29, pp. 419--432,
  2019.

\bibitem{EgilmezPO2018}
Hilmi~E. Egilmez, Eduardo Pavez, and Antonio Ortega,
\newblock ``Graph learning from filtered signals: Graph system and diffusion
  kernel identification,''
\newblock {\em IEEE Transactions on Signal and Information Processing over
  Networks}, vol. 5, no. 2, pp. 360--374, 2018.

\bibitem{PavezOM2017}
Eduardo Pavez, Antonio Ortega, and Debargha Mukherjee,
\newblock ``Learning separable transforms by inverse covariance estimation,''
\newblock in {\em 2017 IEEE International Conference on Image Processing
  (ICIP)}. IEEE, 2017, pp. 285--289.

\bibitem{SimoncelliO2001}
Eero~P Simoncelli and Bruno~A Olshausen,
\newblock ``Natural image statistics and neural representation,''
\newblock {\em Annual review of neuroscience}, vol. 24, no. 1, pp. 1193--1216,
  2001.

\bibitem{IttiKN1998}
Laurent Itti, Christof Koch, and Ernst Niebur,
\newblock ``A model of saliency-based visual attention for rapid scene
  analysis,''
\newblock {\em IEEE Transactions on pattern analysis and machine intelligence},
  vol. 20, no. 11, pp. 1254--1259, 1998.

\bibitem{FracastoroTF2016}
Giulia Fracastoro, Dorina Thanou, and Pascal Frossard,
\newblock ``Graph transform learning for image compression,''
\newblock in {\em 2016 Picture Coding Symposium (PCS)}. IEEE, 2016, pp. 1--5.

\bibitem{LiaoCMYH2018}
Weihang Liao, Gene Cheung, Shogo Muramatsu, Hiroyasu Yasuda, and Kiyoshi
  Hayasaka,
\newblock ``Graph learning \& fast transform coding of {3D} river data,''
\newblock in {\em 2018 Asia-Pacific Signal and Information Processing
  Association Annual Summit and Conference (APSIPA ASC)}. IEEE, 2018, pp.
  1313--1317.

\bibitem{EgilmezCO2020}
Hilmi~E. Egilmez, Yung-Hsuan Chao, and Antonio Ortega,
\newblock ``Graph-based transforms for video coding,''
\newblock {\em IEEE Transactions on Image Processing}, vol. 29, pp. 9330--9344,
  2020.

\bibitem{CLIC2022}
``5th {Challenge on Learned Image Compression dataset},'' Online, June 2022.

\bibitem{GiraultON2018}
Benjamin Girault, Antonio Ortega, and Shrikanth~S. Narayanan,
\newblock ``Irregularity-aware graph {Fourier} transforms,''
\newblock {\em IEEE Transactions on Signal Processing}, vol. 66, no. 21, pp.
  5746--5761, 2018.

\bibitem{EgilmezPO2017}
Hilmi~E. Egilmez, Eduardo Pavez, and Antonio Ortega,
\newblock ``Graph learning from data under {Laplacian} and structural
  constraints,''
\newblock {\em IEEE Journal of Selected Topics in Signal Processing}, vol. 11,
  no. 6, pp. 825--841, 2017.

\bibitem{GershoG2012}
Allen Gersho and Robert~M Gray,
\newblock {\em Vector quantization and signal compression}, vol. 159,
\newblock Springer Science \& Business Media, 2012.

\bibitem{GolubL2013}
Gene~H. Golub and Charles~F. Van~Loan,
\newblock {\em Matrix computations},
\newblock JHU press, 2013.

\bibitem{Bjontegaard2001}
Gisle Bjontegaard,
\newblock ``Calculation of average {PSNR} differences between {RD}-curves,''
\newblock {\em ITU SG16 Doc. VCEG-M33}, 2001.

\bibitem{WaltherK2006}
Dirk Walther and Christof Koch,
\newblock ``Modeling attention to salient proto-objects,''
\newblock {\em Neural networks}, vol. 19, no. 9, pp. 1395--1407, 2006.

\end{thebibliography}

\end{document}